\documentclass[12pt]{article}
\title{Anomalies and divergences in N=4 supergravity}
\usepackage{mathrsfs}
\usepackage{amsmath}

\global\arraycolsep=1pt
\usepackage[colorlinks=true,backref=true,linkcolor=black,anchorcolor=black,citecolor=black,filecolor   =black,menucolor=black,pagecolor=black,urlcolor=black]{hyperref}
\usepackage[Symbol]{upgreek}

\usepackage{amsmath}
\usepackage{amsthm}
\usepackage{amssymb}
\usepackage{mathrsfs}
\usepackage[Symbol]{upgreek}
\usepackage{dsfont}
\usepackage[vcentermath]{youngtab}
\usepackage{textcomp}

\usepackage{latexsym}
\usepackage{graphicx}

\newcommand{\eprint}[1]{{\href{http://arxiv.org/abs/#1}{\texttt{[#1}]}}}
\newcommand{\eprintN}[1]{{\href{http://arxiv.org/abs/#1}{\texttt{#1 [hep-th]}}}}

\makeatletter
\def\section{\@startsection{section}{1}{\z@}{1ex}{1ex}{\bf \large}}
\makeatother

\def\bsh{\backslash}

\newfont{\bbbold}{msbm10 scaled \magstep1}

\def\bbR{\mbox{\bbbold R}}

\def\cA{{\cal A}}

\def\cF{{\cal F}}

\def\cL{{\cal L}}

\def\cN{{\cal N}}
\def\cO{{\cal O}}

\def\cV{{\cal V}}

\newfont{\goth}{eufm10 scaled \magstep1}

\def\a{\alpha}\def\adt{\dot \alpha}
\def\b{\beta}\def\bdt{\dot \beta}
\def\c{\gamma}\def\C{\Gamma}\def\cdt{\dot\gamma}
\def\d{\delta}\def\ddt{\dot\delta}

\def\r{\rho}
\def\s{\sigma}

\def\th{\theta}

\def\be{\begin{equation}}\def\ee{\end{equation}}
\def\bea{\begin{eqnarray}}\def\eea{\end{eqnarray}}
\def\barr{\begin{array}}\def\earr{\end{array}}

\def\xz{\times}

\def\nn{\nonumber}
\def\bd{\begin{document}}
\def\ed{\end{document}}
\def\ba{\begin{array}}
\def\ea{\end{array}}
\def\bea{\begin{eqnarray}}
\def\eea{\end{eqnarray}}
\def\ft#1#2{{\frac{\scriptstyle #1}{\scriptstyle #2}}}
\def\fft#1#2{\frac{#1}{#2}}
\def\sst#1{{\scriptscriptstyle #1}}
\def\oneone{\rlap 1\mkern4mu{\rm l}}

\newcommand{\eq}[1]{(\ref{#1})}
\newcommand{\w}[1]{\\[0.#1cm]}
\def\eqs#1#2{(\ref{#1}-\ref{#2})}
\def\det{{\rm det\,}}
\def\tr{{\rm tr}}

\newcommand{\hoch}[1]{$\, ^{#1}$}
\newcommand{\imperial}{\it\small Theoretical Physics Group, Imperial College London\\ Prince Consort Road, London SW7 2AZ, UK}
\newcommand{\kings}
{\it\small Department of Mathematics, King's College London\\ Strand, London WC2R 2LS, UK}
\newcommand{\uu}
{\it\small Department of Theoretical Physics, Uppsala, Sweden}
\newcommand{\hip}
{\it\small HIP-Helsinki Institute of Physics, P.O. Box 64 FIN-00014
University of Helsinki, Suomi-Finland}
\newcommand{\stock}
{\it\small Department of Theoretical Physics, Stockholm, Sweden}
\newcommand{\cpht}
{\it\small Centre de Physique Th{\'e}orique, Ecole Polytechnique, CNRS\\ 91128 Palaiseau Cedex, France}
\makeatletter

\newcommand{\sa}{/ \hspace{-1.2ex}}
\newcommand{\saa}{/ \hspace{-1.4ex}}
\newcommand{\saaa}{\, / \hspace{-1.6ex}}
\newcommand{\Scal}[1]{\Bigl ({#1} \Bigr )}
\newcommand{\scal}[1]{\bigl ({#1} \bigr )}

\newcommand{\CR}{\nonumber \\*}

\newcommand{\trace}{\hbox {tr}~}
\newcommand{\traceS}{\hbox {tr}_{\scriptscriptstyle \mathfrak{S}}~}

\DeclareMathAlphabet{\mathpzc}{OT1}{pzc}{m}{it}
\def\BRST{\,\mathpzc{s}\,}
\def\aBRST{{\scriptstyle (\mathpzc{s})}}
\def\q{{{\scriptscriptstyle (Q)}}}
\def\qs{{\scriptscriptstyle (Q\mathpzc{s})}}
\def\Qsla{{\mathcal{S}_{\q}}}
\def\Slav{{\mathcal{S}_\aBRST}}
\def\epsilonb{{\overline{\epsilon}}}
\def\bulletup{{\scriptstyle \bullet}}

\newcommand{\gra}[2]{{\scriptscriptstyle (#1 , #2 )}}
\newcommand{\ord}[1]{{\scriptscriptstyle (#1)}}

\def\cL{{\cal L}}
\def\cN{\mathcal{N}}
\def\cO{\mathcal{O}}

\def\ie{{\it i.e.}\ }
\def\eg{{\it e.g.}\ }

\newcommand{\sfrac}[2]{{\scriptstyle \frac{#1}{#2}}}
\newcommand{\stfrac}[2]{{\scriptscriptstyle \frac{#1}{#2}}}

 \def\balpha{{\overline{\alpha}}}
 \def\bbeta{{\overline{\beta}}}
 \def\bgamma{{\overline{\gamma}}}
 \def\bdelta{{\overline{\delta}}}
 \def\bepsilon{{\overline{\epsilon}}}
 \def\bvarepsilon{{\overline{\varepsilon}}}
 \def\bzeta{{\overline{\zeta}}}
 \def\bareta{{\overline{\eta}}}
 \def\btheta{{\overline{\theta}}}
 \def\bvartheta{{\overline{\vartheta}}}
 \def\biota{{\overline{\iota}}}
 \def\bkappa{{\overline{\kappa}}}
 \def\blambda{{\overline{\lambda}}}
 \def\bmu{{\overline{\mu}}}
 \def\bnu{{\overline{\nu}}}
 \def\bxi{{\overline{\xi}}}
 \def\bpi{{\overline{\pi}}}
 \def\brho{{\overline{\rho}}}
 \def\bvarrho{{\overline{\varrho}}}
 \def\bsigma{{\overline{\sigma}}}
 \def\bvarsigma{{\overline{\varsigma}}}
 \def\btau{{\overline{\tau}}}
 \def\bphi{{\overline{\phi}}}
 \def\bvarphi{{\overline{\varphi}}}
 \def\bchi{{\overline{\chi}}}
 \def\bpsi{{\overline{\psi}}}
 \def\bomega{{\overline{\omega}}}

\def\thalf{{\textrm{\tiny\textonehalf}}}
\def\tquarter{{\textrm{\tiny\textonequarter}}}
\def\Ko{{\scriptscriptstyle K}}
\def\tKo{\scriptscriptstyle k }
\def\N{{\mathcal{N}}}
\def\csN{{\fontsize{9.35pt}{9pt}\selectfont \mbox{$\cN$} \fontsize{12.35pt}{12pt}\selectfont }}
\def\cssN{{\fontsize{6.35pt}{6pt}\selectfont \mbox{$\cN$} \fontsize{12.35pt}{12pt}\selectfont }}
\def\csssN{{\fontsize{4.35pt}{4pt}\selectfont \mbox{$\cN$} \fontsize{12.35pt}{12pt}\selectfont }}

\def\ai{{\hat{\imath}}}
\def\aj{{\hat{\jmath}}}
\def\ak{{\hat{k}}}

\newcommand{\red}[1]{ {\color{red} #1 }} 
\newcommand{\blue}[1]{{\color{blue} #1 }}
\newcommand{\green}[1]{{\color{green} #1 }}
\newcommand{\bleu}[1]{ {\color{cyan} #1 }} 

\renewcommand{\thefootnote}{\arabic{footnote}}

\begin{document}

\renewcommand{\thefootnote}{\fnsymbol{footnote}}

\begin{titlepage}
\newcommand{\auth}{\large G.\ Bossard\hoch{1}, P.S.\ Howe\hoch{2} and K.S.\ Stelle\hoch{3}}

\thispagestyle{empty}

\renewcommand{\thefootnote}{\arabic{footnote}}
\vspace{-10mm} \begin{flushright}
{\small CPHT-RR005.1212}\\
{\small KCL-MTH-12-12}\\
{\small Imperial/TP/12/KSS/02}\\
\vskip 1 cm
\end{flushright}

\vspace{10mm} 

\begin{center}
{\Large{\bf  Anomalies and divergences in $\cN=4$ supergravity}}
\vspace{.75cm}

\auth

\vskip 2 em

\begin{itemize}
\item[$^1$]\cpht\item[$^2$] \kings  \item [$^3$] \imperial
\end{itemize}

\end{center}
\vskip 1 em
\begin{abstract}
{The invariants in $D=4, \cN=4$ supergravity are discussed up to the three-loop order (where one expects a general $R^4$ structure). Because there is an anomaly in the rigid $SL(2,\bbR)$ symmetry of this theory, the analysis of possible restrictions on three-loop divergences due to duality needs careful treatment. We show that this anomalous symmetry is still strong enough at the three-loop order to require duality invariance of candidate counterterms. Provided one makes the additional assumption that there exists a full 16-supercharge off-shell formulation of the theory, counterterms at $L\ge2$ loops would also have to be writable as full-superspace integrals. At the three-loop order such a duality-invariant full-superspace integral candidate counterterm exists, but its duality invariance is marginal in the sense that the full-superspace counter-Lagrangian is not itself duality-invariant. We show that, subject to the assumption that a full off-shell quantisation formalism exists, such marginal invariants are not allowable as counterterms. 
}
\end{abstract}



\vspace{1cm}
\hspace{1cm}{\small {\sl bossard@cpht.polytechnique.fr; paul.howe@kcl.ac.uk; k.stelle@imperial.ac.uk}}
\end{titlepage}
\renewcommand{\thefootnote}{\arabic{footnote}}
\setcounter{footnote}{0}

\pagebreak
\setcounter{page}{1}
\section*{Introduction}
Developments in the evaluation of scattering amplitudes using unitarity methods over the past decade or so have made it possible to push the investigation of the onset of ultra-violet divergences in maximal supergravity theories to higher loop orders than would have been possible using conventional Feynman-diagram techniques. In particular, it has been shown that $D=4, \cN=8$ supergravity is finite at three loops ($R^4$) \cite{Bern:2007hh}, and that $D=5$ maximal supergravity is finite at four loops ($\partial^6 R^4$) \cite{Bern:2009kd}, despite the existence of corresponding counterterms, at least at the linearised level \cite{Kallosh:1980fi,Howe:1981xy,Bossard:2009sy}. Since these invariants are F-type, \ie correspond to integrals over fewer than the maximal number of odd superspace coordinates, it might have been thought that they should be protected by superspace non-renormalisation theorems \cite{Bossard:2009sy}, but it is difficult to justify this argument because there are no known off-shell versions of maximal supergravity that realise all of the supersymmetries linearly. Indeed, such off-shell versions cannot exist in every dimension because it is known that divergences do occur for F-type counterterms in $D=6$ and $D=7$ above one loop \cite{Bern:1998ug}. However, these finiteness results can be explained instead by duality-based arguments. $E_{7(7)}$ Ward identities can be defined at the cost of manifest Lorentz covariance \cite{Hillmann:2009zf,Bossard:2010dq}, and can be shown to be non-anomalous.\footnote{The absence of a supersymmetric anomaly for the $E_{7(7)}$ Ward identities that cannot be removed by supersymmetric non-invariant counterterms has not been rigorously established at all orders in perturbations theory. Nonetheless, the complete characterisation of the supersymmetry invariants of type $R^4$, $\partial^4 R^4$, $\partial^6 R^4$, $\partial^8 R^4$   
\cite{Drummond:2003ex,Elvang:2010kc,Bossard:2011tq} allows one to prove that such an anomaly cannot appear before eight loops.} These Ward identities imply that the counterterms associated to logarithmic divergences must be $E_{7(7)}$ invariant. The unique $SU(8)$ invariant $R^4$ candidate counterterm can be proved to violate $E_{7(7)}$ symmetry from a perturbative scattering amplitude approach \cite{Elvang:2010kc} and from a direct field-theoretic argument \cite{Bossard:2010bd} making use of dimensional reduction and of the uniqueness of the $D=4$ counterterms at the linearised level \cite{Drummond:2003ex}.  In addition, there is no superspace measure for the $R^4$ invariant at the full non-linear level, while an analysis of the closed super-four-form that does define the supersymmetric invariant leads to the same conclusion: there is no three-loop acceptable counterterm that is both $\cN=8$ supersymmetric and $E_{7(7)}$ duality-invariant \cite{Bossard:2010bd}. Furthermore, these arguments can be extended to the other two F-term invariants in $D=4$ arising at the five and six-loop orders \cite{Bossard:2010bd,Beisert:2010jx}, there being no four-loop invariant \cite{Drummond:2003ex}. One can then use dimensional reduction and the known divergences at one, two and three loops in $D=8,7$ and 6, respectively,  to show that these are the only F-term divergences that can arise in maximal supergravity in any dimension. This result can also be seen  from an analysis of the conjectured duality properties of superstring theory \cite{Green:2010sp,Green:2010kv}. It has also been suggested that $E_{7(7)}$ symmetry could be even more restrictive and that $\cN=8$ supergravity might, as a consequence, be finite at all orders \cite{Kallosh:2011dp,Kallosh:2011qt}.

It therefore seems that maximal supergravity must be ultra-violet finite through at least six loops in $D=4$, and that there are no divergences that correspond to the known linearised BPS counterterms (F-terms) \cite{Kallosh:1980fi,Howe:1981xy,Drummond:2003ex,Elvang:2010jv,Drummond:2010fp}. At the seven-loop order, we reach the borderline between F-term and D-term invariants. At this order, there would seem to be a candidate D-term invariant, the volume of superspace, which is manifestly symmetric with respect to all symmetries and which would be difficult to protect by conventional field-theoretic non-renormalisation arguments. However, it is now known that the volume of superspace vanishes on-shell for any $\cN$ in $D=4$ although there is still an $\cN=8$ seven-loop invariant that can be written as a manifestly duality-invariant harmonic-superspace integral over $28$ odd coordinates \cite{Bossard:2011tq}. The situation at this order is therefore somewhat ambiguous, although it is unlikely that there is an off-shell formulation of the maximal supergravity theory preserving all the supersymmetries linearly which could be used to try to justify the absence of a seven-loop divergence. A direct computational resolution of this ambiguity would seem to be a tall order, at least in the near future, but a similar situation arises in the half-maximal case which is more tractable from both computational and formal points of view.

In $D=4,\cN=4$ supergravity the F/D borderline occurs at the three-loop level, \ie for $R^4$ type counterterms. It has recently been shown that half-maximal supergravity is finite at this order \cite{Bern:2012cd,Tourkine:2012ip,Tourkine:2012vx} and that this state of affairs persists in  $D=5$ \cite{Bern:2012gh} (where the relevant loop order is two) and in the presence of vector multiplets \cite{Tourkine:2012ip,Tourkine:2012vx}. These finiteness results have been obtained from scattering-amplitude computations \cite{Bern:2012cd,Bern:2012gh} and from string theory \cite{Tourkine:2012ip,Tourkine:2012vx}. Field-theoretic arguments in support of these results have also been given using 
duality and conformal symmetry \cite{Kallosh:2012ei,Ferrara:2012ui}. From the counterterm point of view, the situation resembles seven loops in $\cN=8$ because the natural candidate for the $R^4$ invariant would be the volume of superspace. As in $\cN=8$, this vanishes in $\cN=4$ \cite{Bossard:2011tq}, but one can also construct $R^4$ invariants as harmonic-superspace integrals over twelve odd coordinates instead of the full sixteen. As we shall show, the unique duality-invariant  counterterm of this type can be re-expressed as a full-superspace integral with an integrand that is not itself duality-invariant; in fact it is the K\"ahler potential of the scalar manifold. The issue is therefore to understand if this property is enough to rule out this counterterm as a possible divergence.

For $\cN=4,D=4$ supergravity, the duality group is $SL(2,\bbR)$. This symmetry is anomalous \cite{Marcus:1985yy}, but we shall show that the anomalous Ward identities still require the three-loop counterterm to be duality-invariant.  To show that such an invariant does not correspond to a divergence, we shall need to assume that there is an off-shell formulation of the theory that preserves all of the supersymmetries linearly as well as duality symmetry. This is not possible in conventional superspace for the case in hand \cite{Rivelles:1982gn}, so we shall have to make use of harmonic-superspace techniques \cite{Galperin:1984av,Karlhede:1984vr}. In addition,  duality symmetry is not compatible with manifest Lorentz invariance, and so it seems likely that  some version of the light-cone harmonic-superspace formalism will be required \cite{Sokatchev:1985tc}.

To summarise, the dimension-eight ($R^4$) invariant in $\cN=4,D=4$ supergravity can be considered to be on the F/D borderline because it can be expressed either as an integral of a duality-invariant integrand over twelve odd coordinates or as a full-superspace integral whose integrand is not itself invariant even though the full integral is. If we make the assumption that there exists  an off-shell version of the theory that preserves all of the supersymmetries linearly as well as the duality symmetries, then, as we shall show, the divergences would have to correspond to full-superspace integrals with integrands that are duality-invariant and hence would be absent at this loop order. 

One clear result emerging from our analysis is the fact that the uniqueness of the dimension-eight invariant means that the absence of divergences in four-point amplitudes implies that they will also be absent in all higher-point ones. The borderline F/D problem is difficult to analyse from our field-theoretic point of view, but for higher loops there will certainly be candidate counterterms that are purely D-type and whose integrands are invariant with respect to all known symmetries. In this sense, an unambiguous test of ``miraculous'' ultra-violet cancellations in half-maximal supergravity requires calculations at one loop higher than those that have been carried out to date.
\section*{$D=4,\N=4$ supergravity}

  The fields for $D=4,\cN=4$ supergravity consist of two scalars parametrising the coset $U(1)\bsh SL(2,\bbR)$, a quadruplet of Weyl spinors $\chi_\a^i$, transforming under the fundamental representation of $SU(4)$, six vector fields with field strengths $M_{\a\b ij}$, four gravitinos with field strengths $\r_{\a\b\c i}$, and the graviton with on-shell field-strength the Weyl tensor $C_{\a\b\c\d}$, together with their complex conjugates. Here, $\a,\b, ...$ are two-component spinor indices. In the linearised case these fields fit into a chiral superfield $W$. In the non-linear case they appear in the various components of the superspace tensors: the torsion, the curvature and the field strength of the vectors \cite{Howe:1981gz}. The dimension-zero torsion remains the same as in the flat case and is given by
\be 
T_\alpha^i{}_{\dot{\beta} j}{}^c = - i \delta^i_j \sigma^c_{\alpha\dot{\beta}}\ , \ee
while the spinor field appears at dimension one-half,
\be T_{\alpha}^i{}_\beta^j{}^{\dot{\gamma} k} = - \varepsilon_{\alpha\beta} \varepsilon^{ijkl} \bar\chi^{\dot{\gamma}}_l \ .
\ee
The scalars are described by an $SU(1,1)\ (\cong SL(2,\bbR))$ matrix $\cV$ which can be parametrised in the form
\be
\cV=\left(\barr{cc} U & UT \cr \bar U\bar T & \bar U\earr\right)
\ee
where 
\be U \bar U ( 1 - T \bar T ) = 1 \ .\ee
The $U(1)$ gauge-invariant complex scalar superfield $T$ can be considered to be a coordinate on the unit disc. Both $U$ and $T$ are chiral while 
\be D_\alpha^i U = \frac{1}{U} \frac{ \bar T }{1 - T \bar T } \chi_\alpha^i \ , \qquad D_\alpha^i T = \frac{1}{U^2} \chi_\alpha^i \ .\ee

\section*{$R^2$ invariants}
\label{Rdeux} 
The first possible on-shell invariants in pure $\cN=4$ supergravity are all of generic $R^2$ structure. Although the relevant $SL(2,\bbR)$ invariants vanish on-shell for trivial spacetime topology, there do exist non-vanishing invariants with a non-trivial dependence on the complex scalar field, which play a key r\^ole in the discussion of anomalies. 

Because both $\bar U$ and $\bar  T$ are antichiral, one might think that one could define $R^2$ type invariants as anti-chiral superspace integrals of functions of these fields.
However, owing to the presence of the dimension one-half torsion component, there is no chiral measure, as in the case of the $R^4$ invariant in IIB supergravity \cite{Howe:1983sra,Skenderis}. Instead we can construct $R^2$ invariants involving an arbitrary anti-holomorphic function $\cF(\bar T)$ from closed superforms. 

Any invariant in $D=4$, whether or not it is expressible  as a superspace integral, can be written in terms of a closed  super-four-form $\cL$ which can be expanded as $\cL=\sum_{p+q=4} \cL_{p,q}$ where $(p,q)$ denote the even (odd) degrees of the components of the form in a preferred basis, see \eg \cite{Gates:1997ag}. The invariant is then given by integrating $\iota^*\cL$, evaluated at $\theta=0$, over spacetime, where $\iota$ injects spacetime into superspace. This pull-back involves the sum of the components $\cL_{p,q}$ contracted with $p$ spacetime vielbein forms and $q$ spacetime gravitino forms.  The differential decomposes into four components \cite{Bonora:1986ix},
\be d =d_0 + d_1 + t_0  +t_1\ , \ee
of bi-degrees  $(1,0)$, $(0,1)$, $(-1,2)$ and $(2,-1)$ respectively. The first two components of $d$ are respectively even and odd differential operators while the other two are algebraic involving the dimension-zero and dimension-three-halves torsions. The latter is not relevant to the present discussion, while the former can be defined in terms of the contraction operator $ \iota_c   E^a \equiv \delta^a_c$  
and  the dimension-zero torsion by
\be t_0 L_{p,q} = -iE^{\bdt i} \wedge E^\alpha_i \, (\s^c)_{\alpha\bdt } \wedge  \iota_c L_{p,q} \ .\ee
Since $d^2=0$ it follows that $t_0$ is nilpotent so that we can define cohomology groups $H_t^{p,q}$ \cite{Bonora:1986ix}. This means that we can analyse superspace cohomology in terms of elements of this group. For our case it is easy to see that $H_t^{p,q}=0$ for $p>0$ so that any closed four-form is generated by a non-exact $(0,4)$-form that is $d_1$-closed in $t_0$-cohomology, \ie an element in spinorial cohomology \cite{Cederwall:2001dx}. In four dimensions, we can split the odd indices into dotted and undotted ones, so that $\cL_{0,4}=\sum_{r+s=4} M_{0,r,s}$. 

The closed, complex four-form $R_{\adt\bdt}\wedge R^{\adt\bdt}$  gives rise to a trivial cocycle in a topologically trivial spacetime. Nevertheless, it can be used as a starting point from which to construct the invariants we are interested in. From the fact that $R_{\adt i \bdt j,\cdt\ddt}=0$ we can see that $\cL_{0,4}=M_{0,4,0} + M_{0,3,1}$, up to $t_0$ exact terms, whereas a cocycle for a chiral integral would also have $M_{0,3,1}=0$. One can then show after some algebra that the following expressions define closed super-four-forms $\cL[\cF]$ for any anti-holomorphic function $\cF(\bar T)$:
\begin{multline} \label{R2Cocycle} M_{0,4,0}\rightarrow M_\a^i{}_\b^j{}_\c^k{}_\d^l  = \varepsilon_{\alpha\beta} \varepsilon_{\gamma\delta} \Bigl( \cF^\ord{0}(\bar T) M_{\dot{\alpha}\dot{\beta}}^{ij} M^{\dot{\alpha}\dot{\beta} kl} - \bar U^{-2} \bar \partial \cF^\ord{0}(\bar T) \varepsilon^{ijpq} \chi_{\dot{\alpha} p} \chi_{\dot{\beta} q} M^{\dot{\alpha}\dot{\beta} kl}  \Bigr . \\ \Bigl .+ \frac{1}{6} \bar U^{-4} \scal{ \bar \partial - \frac{2 T}{ 1 - T \bar T} } \bar \partial  \cF^\ord{0}(\bar T) \varepsilon^{ijpq} \varepsilon^{klrs} \chi_{\dot{\alpha} p} \chi_{\dot{\beta} q} \chi_r^{\dot{\alpha}} \chi_s^{\dot{\beta}} \Bigr) \ +   \circlearrowleft  \end{multline} 
and
\be M_{0,3,1}\rightarrow  M_\a^i{}_\b^j{}_\c^k{}_{\ddt l}  = - \varepsilon_{\alpha\beta} \varepsilon^{\dot{\eta}\dot{\varsigma}}  \chi_\gamma^k \chi_{\dot{\eta} l} \Bigl( \cF^\ord{0}(\bar T) M_{\dot{\delta}\dot{\varsigma}}^{ij} -\frac{1}{3}  \bar U^{-2} \bar \partial \cF^\ord{0}(\bar T) \varepsilon^{ijpq} \chi_{\dot{\delta} p} \chi_{\dot{\varsigma} q}  \Bigr)  \ +   \circlearrowleft  \ , \ee
where $ \circlearrowleft$ denotes permutations to be added to obtain the right symmetry structure. 

By construction, for $\cF = 1$, the  chiral superform reduces to  $R_{\dot{\alpha}\dot{\beta}} \wedge R^{\dot{\alpha}\dot{\beta}}$ so that the general  invariant of this type includes the term 
\be \cL[\cF] = \cF(\bar T) R_{\dot{\alpha}\dot{\beta}} \wedge R^{\dot{\alpha}\dot{\beta}} + \dots \ \ .\ee
This invariant is of course complex, and the associated real invariants will be obtained from its real and imaginary parts, which are respectively even and odd with respect to parity. 

\section*{$R^4$ type invariants} 
It is well-known that there are no $R^3$ type invariants in supergravity, so that  the next non-trivial invariants are of $R^4$  type which arise at three loops in $D=4$. At this order, in $\cN=4$, examples of such invariants are given by full-superspace integrals of arbitrary functions of the complex scalar superfield $T$. To analyse these, it will turn out to be useful to use harmonic-superspace techniques in a supergravity context.

In flat superspace we recall that a G-analytic (G for Grassmann) structure of type $(p,q)$ consists of a set of $p\ D$s and $q\ \bar D$s that mutually anti-commute, and that such sets can be parametrised by the coset spaces $(U(p)\xz U(\cN-(p+q))\xz U(q))\bsh U(\cN)$, which are compact, complex manifolds (flag manifolds). Harmonic superspaces consist of ordinary superspaces augmented by the above cosets. However, in curved superspace one has to check that these derivatives, suitably extended to include the harmonic directions, remain involutive in the presence of the non-trivial geometry. It turns out that this is only possible when both $p,q\leq 1 $ for $\cN>4$, but for $\cN=4$ one can have $p,q\leq 2 $ \cite{Hartwell:1994rp}.

Let us consider first the $\cN=4$ theory in $(4,1,1)$ harmonic superspace. The harmonic variables, $u^1{}_i, u^r{}_i, u^4{}_i$ (and their inverses $u^i{}_1, u^i{}_r, u^i{}_4$), where $r=2,3$, can be used to parametrise the coset $\scal{U(1) \times U(2) \times U(1)}  \setminus U(4)$ in an equivariant fashion. We can associate four odd normal coordinates with the four involutive odd directions, as in  \cite{Bossard:2011tq,KuzenkoRY}, and use these to relate full-superspace integrals to integrals over the remaining twelve odd coordinates, \ie over $(4,1,1)$ analytic superspace (the analogue of chiral superspace for G-analyticity). This programme was carried out in pure $\cN=4$ and $\cN=8$ supergravities in \cite{Bossard:2011tq} to show that the full-superspace integral of the Berezinian (superdeterminant) of the supervielbein, $E$, vanishes subject to the classical equations of motion. We refer to  \cite{Bossard:2011tq,KuzenkoRY} for more details; here, we will simply state that the full-superspace integral of an arbitrary function $H(T,\bar T)$ can be rewritten as
\bea  &&  \int d^4 x d^{16} \theta \, E \, H(T,\bar T) = \frac{1}{4} \int d\mu_{\scriptscriptstyle (4,1,1)} \varepsilon^{\alpha\beta} \varepsilon^{\dot{\alpha} \dot{\beta}}  D_\alpha^1 D_\beta^1  D_{\dot{\alpha}4} D_{\dot{\beta}4}  \, H(T,\bar T) \CR
&=&  \frac{1}{4} \int d\mu_{\scriptscriptstyle (4,1,1)} \varepsilon^{\alpha\beta} \varepsilon^{\dot{\alpha} \dot{\beta}}  \chi_\alpha^1 \chi_\beta^1  \chi_{\dot{\alpha}4} \chi_{\dot{\beta}4} \, \scal{ ( 1- T \bar T )^2 \partial \bar \partial - 2 }  ( 1 - T \bar T )^2 \partial \bar \partial \, H(T,\bar T) \ ,\eea
where numerical indices are obtained from $SU(4)$ indices by contracting with the appropriate $u$, \eg $D_\a^1= u^1{}_i D_\a^i$.  Clearly this integral vanishes if $H$ is an eigenfunction of the scalar target-space Laplace operator with eigenvalue $0$ or $2$, and in particular if $H(T,\bar T)$ is a constant, or more generally a holomorphic function. However, one also straightforwardly computes that
\be \scal{ ( 1- T \bar T )^2 \partial \bar \partial - 2 }  ( 1 - T \bar T )^2 \partial \bar \partial \, \Scal{ - \mbox{ln}\left({1-T \bar T} \right) }= 1 \ee
and therefore the full-superspace integral of $\mbox{ln}\left({1-T \bar T}\right)$ is duality-invariant. This is the K\"{a}hler potential of the symmetric space $SU(1,1)/U(1)$, which in terms of $\uptau = i \frac{1-T}{1+T}$ is $K= - \mbox{ln}( \mbox{Im}[\uptau])$. Under a duality transformation this transforms into the sum of a holomorphic function of $\uptau$ and its conjugate so that its integral over superspace is invariant. 

For any function $G(T,\bar T)$ we can define a $(4,1,1)$ analytic superspace integral by
\be  S[G]=  \frac{1}{4} \int d\mu_{\scriptscriptstyle (4,1,1)} \varepsilon^{\alpha\beta} \varepsilon^{\dot{\alpha} \dot{\beta}}  \chi_\alpha^1 \chi_\beta^1  \chi_{\dot{\alpha}4} \chi_{\dot{\beta}4}  \, G(T,\bar T) \label{R4Pure} \ee
because the integrand is G-analytic, \ie annihilated by $D_\a^1$ and $D_{\adt 4}$. This follows from the properties of $\chi$ under differentiation and from the fact that $D_\a^1 T\propto \chi_\a^1$, so that differentiating $G$ leads to cubic (and hence vanishing) expressions in $\chi^1$ or $\bar\chi_4$. This class of invariants reproduces all the possible invariants at this dimension in the linearised approximation
\bea  \int d\mu_{\scriptscriptstyle (4,1,1)} \varepsilon^{\alpha\beta} \varepsilon^{\dot{\alpha} \dot{\beta}}  \chi_\alpha^1 \chi_\beta^1  \chi_{\dot{\alpha}4} \chi_{\dot{\beta}4} G(T,\bar T) &\sim&    \int d^4 xd^{16} \theta\,  W^2 \bar W^2 G(W,\bar W) \\
 &\sim&  \int d^4 x\,  G(t,\bar t) C_{\alpha\beta\gamma\delta} C^{\alpha\beta\gamma\delta} C_{\dot{\alpha}\dot{\beta}\dot{\gamma}\dot{\delta}} C^{\dot{\alpha}\dot{\beta}\dot{\gamma}\dot{\delta}}+ \dots \ ,\nn \eea
where $t=T|_{\th=0}$, and therefore includes all the $R^4$ type invariants. Moreover, a generic invariant in this class can always be rewritten as a full-superspace integral of a function $H$ which is a solution of the equation
\be \scal{ ( 1- T \bar T )^2 \partial \bar \partial - 2 }  ( 1 - T \bar T )^2 \partial \bar \partial \, H(T,\bar T) = G(T,\bar T) \ . \ee

\section*{The $\mathfrak{sl}_2(\mathds{R})$ anomaly and the three-loop divergence}

The $SL(2,\mathds{R})$ duality group acts on the complex  scalar superfield 
\be  \uptau[ T] \equiv i \frac{1 -T }{1 + T } = a + i e^{-2\phi} + \mathcal{O}( \theta) \ee
in the standard way. In order to discuss its anomaly, it is convenient to introduce anticommuting parameters for an infinitesimal (BRST) 
$\mathfrak{sl}_2\mathds{R}$ transformation as
\be \left( \begin{array}{cc} h & \ \ e \\ f &\  -h \end{array} \right) \in \mathfrak{sl}_2 \ .\ee  
One defines the BRST-like operator associated to the $\mathfrak{sl}_2\mathds{R}$ symmetry by
\be \delta  \uptau   = e + 2 h \uptau - f \uptau^2\ ,  \qquad \delta f = - 2 h f\ ,  \quad \delta h = e f \ , \quad \delta e = 2 h e \ . \ee 
One then straightforwardly checks that $f \uptau - h$ is a representative of the unique cohomology class of $\delta$ linear in the anticommuting parameters. Although this term is complex, its real part is $\delta$-exact, \ie $f ( \uptau + \bar \uptau ) - 2 h = - \delta\,  \mbox{ln}( \uptau - \bar \uptau )$. 

This symmetry is non-linear and is only defined by Slavnov--Taylor functional identities. We shall discuss this in detail in an accompanying paper, but here we content ourselves with  a somewhat informal discussion which does, nevertheless, lead to the right answer. There are no one-loop divergences in the theory, but the one-loop effective action $\C^{(1)}$ is nonetheless anomalous, $\d \C^{(1)}\sim\cA^\ord{1}$.

Because the invariant $\iota^* \cL [\cF]$ defined above is linear in the function $\cF$, one finds that the anomaly functional is 
\bea \cA^\ord{1} &=& \frac{i}{16\pi^2}   \int  \, \iota^* \scal{ \cL[ f \bar \uptau -h] - \bar \cL[f \uptau - h ]}   \\
&=& \frac{1}{16\pi^2}  \int  \scal{ f e^{-2\phi} R_{ab} \wedge R^{ab} + ( f a - h ) \frac{1}{2} \varepsilon_{abcd} R^{ab} \wedge R^{cd} + \dots } \ . \label{AnomalyInvariant} \nn
 \eea
 Indeed, the anomalous Ward identity implies that the variation of the 1PI generating functional produces a cohomologically non-trivial term $ f e^{-2\phi} R_{ab} \wedge R^{ab}$ \cite{Bossard:2010dq}, but consistency with supersymmetry then implies that this must occur together with the cohomologically trivial term $( f a - h ) \frac{1}{2} \varepsilon_{abcd} R^{ab} \wedge R^{cd}$. This term implies that there is a current anomaly in the scaling symmetry corresponding to the parameter $h$.

There is no candidate for a two-loop anomaly so we deduce that the variation of the divergent part of the three-loop effective action is

\be
\d \C^{(3)}_{\rm div} \sim [\cA^{(1)}\cdot \C]^{(2)}_{\rm div}
\ee
where the right-hand side denotes the divergent part of the two-loop effective action with one insertion of the one-loop anomaly.  The $h$-dependent part of the anomaly is a total derivative and hence does not contribute, so that the non-invariance of the effective action at three loops is proportional to the parameter $f$; in other words, the parabolic sub-algebra determined by $e,h$ remains unbroken. 

A general supersymmetric three-loop invariant can be written as a harmonic-superspace integral $S[G]$ as in \eq{R4Pure}. 
Because $\delta S[G^\ord{3}] = S[\delta G^\ord{3}]$ and  $S[\delta G^\ord{3}]$ is non-zero for any non-zero function $\delta G^\ord{3}$, the function $G^\ord{3}(\uptau, \bar \uptau)$ must itself be invariant with respect to the action of the parabolic subgroup, \ie 
\be  ( \partial + \bar \partial ) G^\ord{3}(\uptau ,\bar \uptau) = 0 \ , \qquad ( \uptau \partial + \bar \uptau \bar \partial )   G^\ord{3}(\uptau ,\bar \uptau)= 0\ .   \ee
One can easily check that the only solution is a constant. At this order, therefore, shift and scaling invariance together with local supersymmetry are enough to require that the invariant be fully duality-invariant,\footnote{Note that in components, $\tau = a + i e^{-2\phi}$, these equations just imply  that the invariant can only depend on the scalar through contractions with the vector fields and $e^{2\phi} \partial_\mu a$ and $\partial_\mu \phi$, which are not necessarily duality-invariant.} and, as a corollary, the anomaly operator cannot get renormalised at the two-loop order. The only available counterterm consistent with all symmetries at this order is the $R^4$ type duality-invariant 
\be S_3 =  \int d\mu_{\scriptscriptstyle (4,1,1)} \varepsilon^{\alpha\beta} \varepsilon^{\dot{\alpha} \dot{\beta}}  \chi_\alpha^1 \chi_\beta^1  \chi_{\dot{\alpha}4} \chi_{\dot{\beta}4} \ . \ee
This invariant can also be written as the full-superspace integral of the K\"{a}hler potential, as we have seen previously. 

\section*{Algebraic renormalisation in superspace}
We shall now argue that such a three-loop divergence is not allowed if we assume that there is an off-shell version of theory that preserves all of the supersymmetries linearly as well as duality. Standard non-renormalisation theorems in superspace then imply that any acceptable counterterm should be a full-superspace integral of the background fields that does not depend explicitly on the quantum prepotentials (which may have low dimensionality).\footnote{See, e.g. \cite{Bossard:2009sy} for a review of this topic.} In $\cN=4$ supergravity at three loops there is such a counterterm, namely the full-superspace integral of the K\"ahler potential of the scalar manifold, but it has the property that the integrand is not itself duality-invariant. We shall now argue that this  is enough to forbid the occurrence of $R^4$ divergences.

The Callan--Symanzik equation implies that an $R^4$ three-loop divergence necessarily would mean that the Lagrangian density considered as a local operator insertion also would be renormalised into the corresponding $R^4$ density  \cite{PS,Sorella1,Sorella2}. Within an off-shell formulation of the theory in superspace, this divergence would be associated to a three-loop renormalisation of the Lagrange density in superspace, which would necessarily depend on prepotentials. Although we do not know  this formulation explicitly, supersymmetric gauge theories formulated in superspace within the background field method generically admit a Lagrangian density that does not depend on the background prepotentials explicitly (up to a purely classical term that does not affect the Feynman rules in superspace). We shall assume that this density transforms consistently with respect to duality transformations in such a way that one can define $SL(2,\mathds{R})$ Ward identities. This implies that the variation of the Lagrange density is the total derivative of a vector density 
\be \delta \cL^\ord{0} = (-1)^M \partial_M   \cL^{\ord{0} \, M} \  . \ee
We will refer to this vector density as a co-form of degree one, and note that consistency requires the existence of a chain of co-forms satisfying
\bea 
 \delta \cL^{\ord{0} \, M}  &=&(-1)^{N}  \partial_N   \cL^{\ord{0} \, NM} \ , \CR
 \delta \cL^{\ord{0} \, NM}  &=&(-1)^P  \partial_P   \cL^{\ord{0} \, PNM} \ , \CR
 \delta \cL^{\ord{0} \, PNM} &=& 0  \ . 
 \label{DualityDescent} \eea
where we define a co-form of degree $n$ as an object transforming as the tensor product of a density with the graded antisymmetric tensor product of $n$ vectors. Of course, on a Riemannian $d$-dimensional manifold such an object would be equivalent to a $(d-n)$-form via contraction with the Levi--Civita tensor, but in superspace they are distinct objects. Note that a co-form or a form can have an arbitrarily high degree in superspace, and there is correspondingly no notion of a top form; however, there are only three anti-commuting parameters associated to $\mathfrak{sl}_2$, and therefore the last co-form in the above sequence will have (at most)  degree three ($\propto e f h $). 

One can ensure duality invariance by introducing a source for each of these co-forms,
\be \int d^4x d^{16} \theta  \Scal{  \cL^\ord{0} u + \cL^{\ord{0} \, M}  u_M + \tfrac{1}{2} \cL^{\ord{0} \, NM} u_{MN}   + \frac{1}{6} \cL^{\ord{0} \, PNM} u_{MNP}   }\ , \ee
such that the sources transform with respect to duality as an extended cocycle 
\be ( d + \delta  ) \Scal{ u + dz^M u_M + \frac{1}{2} dz^N \wedge dz^M u_{MN} + \frac{1}{6}dz^P \wedge  dz^N \wedge dz^M u_{MNP}  } = 0 \ . \ee  
The extended cocycle is a cohomology class of the extended exterior derivative $d + \delta $, so one can consider the chain of co-forms as defining a cohomology class. Since a density Lagrangian $\cL$ does not depend on anticommuting duality parameters, it accordingly cannot be $\delta$-exact. However, a Lagrange density is only defined up to a total divergence, and therefore the whole chain of co-forms is itself only defined up to a extended co-form trivial in cohomology.  As long as the duality Ward identities are satisfied, the whole chain of co-forms must be renormalised consistently as a single cohomology class. 

This construction is a superspace generalisation of the one developed in \cite{PS,Sorella1,Sorella2} in the framework of algebraic renormalisation. 

Assuming the existence of a duality-invariant formulation of the theory in superspace, the co-forms associated to the classical Lagrangian density are also expected not to  depend on the background-field prepotentials. It then follows that the chain of co-forms associated to a duality-invariant candidate counterterm must also be expressible in terms of the potentials (and no explicit prepotentials). 

The unique duality-invariant counterterm  that can be written as a full-superspace integral is the integral of the K\"{a}hler potential. The associated density is not duality-invariant, but satisfies instead 
\be \delta \Scal{ -E\, \mbox{ln}\Scal{ \tfrac{-i}{2} ( \uptau - \bar \uptau )}} = - 2 h \,E  + f \,E ( \uptau + \bar\uptau ) \ . \label{densityvar}\ee
One cannot rely on the variation in $f$ because of the anomaly, but the variation in $h$ should be expressible as the divergence of a quantity that does not depend on prepotentials in order for the counterterm to be allowed. 
However, the scalar field $\uptau$ and the Berezinian of the supervielbein on the right-hand side of \eqref{densityvar} can only be expressed as total derivatives of functions depending explicitly on the hypothetical prepotentials of the theory.   

This argument remains rather formal in the absence of an explicit formulation of the theory in superspace, and we shall discuss the example of a two-dimensional non-linear sigma model over a symmetric K\"{a}hler space with $(2,2)$ supersymmetry (and no torsion) in an accompanying paper. In this case one can work out the algebraic renormalisation proof in full detail, and confirm that the beta function is one-loop exact within the background field method \cite{Howe:1986ys}.\footnote{In this paper it was shown that the possible logarithmic divergences beyond one-loop must be associated to the superspace integral of functions of the target space Riemann tensor. For a symmetric space, such functions are constant and integrate to zero.} 
\section*{Conclusions}

In this paper we have discussed the possible ultra-violet divergences that can arise in $D=4, \cN=4$ supergravity at three loops. We have argued, provided that  some assumptions regarding off-shell formalisms are made, that this theory should indeed be finite at this order. The key observation is that, although the candidate counterterms seem superficially to be D-terms, \ie integrals over the full sixteen-theta superspace, the fact that the volume of superspace vanishes means that there are no candidate counterterms with manifestly duality-invariant full-superspace integrands. Instead, the relevant duality-invariant integrals can either be written as full-superspace integrals of integrands that are not themselves invariant, or as sub-superspace integrals of invariant integrands, \ie F-terms. We have called this situation the F/D borderline since the status of these invariants is ambiguous. Given the existence of suitable off-shell versions of the theories that preserve all of the supersymmetries linearly, as well as duality, we have argued that the F-term character wins out and that these invariants are therefore protected.

Similar arguments can be applied to $D=5$ supergravity (at two loops) and to half-maximal supergravities coupled to vector multiplets. In the former case, there is no anomaly and the duality symmetry simply consists of constant shifts of the dilaton. Since the volume of superspace vanishes for the $D=5$ half-maximal case, it follows that a suitable full-superspace counterterm is the integral of the dilaton for which the integrand is obviously not shift-invariant. Another simplification in $D=5$ is that the preservation of duality symmetry does not require one to relinquish manifest Lorentz invariance. The addition of vector multiplets, however, does cause more problems, particularly in $D=4$. In their presence, there is a one-loop divergence, together with more possibilities for anomalies and the discussion of the invariants is more involved due to mixing between the gravitational and matter sectors.  All of these matters will be discussed elsewhere.

For the future, it would clearly be of interest to construct the off-shell formalism whose existence we have relied upon in our arguments, although this is not an easy problem. 
Whether or not this can be done successfully, it seems difficult to imagine any purely field-theoretic argument that could protect yet higher-loop counterterms against ultra-violet divergences.\footnote{R. Kallosh has argued that duality might also be effective in protecting D-terms.} This is because there are no obstructions to the construction of counterterms that are full-superspace integrals that are manifestly invariant under all symmetries. This being the case, there is an obvious challenge on the computational side. If it turns out that, \eg, $\cN=4, D=4$ supergravity is finite at four loops, then all bets would be off regarding the perturbative finiteness of $\cN=8$ supergravity.

 \end{document}